\documentclass[a4paper]{article}
\usepackage{epsfig}
\begin{document}
\newcommand{\be}{\begin{equation}}
\newcommand{\ee}{\end{equation}}
\newcommand{\bea}{\begin{eqnarray}}
\newcommand{\eea}{\end{eqnarray}}
\newcommand{\nn}{\nonumber \\}
\newcommand{\de}{{\rm d}}
\newfont{\Kapfont}{cmbx10 scaled 1728}

\vspace*{1cm}
\begin{center}
{\Kapfont Floating Bodies of Equilibrium II}
\end{center}
\vspace{1cm}

\begin{center}
\bf Franz Wegner, Institut f\"ur Theoretische Physik \\
Ruprecht-Karls-Universit\"at Heidelberg \\
Philosophenweg 19, D-69120 Heidelberg \\
Email: wegner@tphys.uni-heidelberg.de
\end{center}
\vspace{1cm}

\paragraph*{Abstract}In a previous paper (physics/0203061) ''Floating bodies of
equilibrium I'' I have shown that there exist two-dimensional
non-circular cross-sections of bodies of homogeous density $\rho\not=1/2$ which
can float in any orientation in water, which have a $p$-fold rotation axis.
For given $p$ they exist for $p-2$ different densities. However, this was
found only in a Taylor-expansion in a parameter which described the distortion
from the circular shape up to seventh order. Here a differential equation for
the boundary curve is given and the non-circular boundary curve can be
expressed in terms of an elliptic integral.

\section{Introduction}

Stanislaw Ulam asks in the Scottish Book \cite{Scottish} (problem 19),
whether a sphere is the only solid of uniform density which will float in
water in any position. In a recent paper\cite{Wegner} I considered the
two-dimensional problem for $\rho\not=1/2$. (The case $\rho=1/2$ has been
solved by Auerbach \cite{Auerbach}). I
was able to obtain non-circular two-dimensional cross-sections of bodies which
can flow in any orientation in water by a perturbative expansion around the
circular solution. These cross-sections have a $p$-fold rotation symmetry.
In polar coordinates $r$ and $\psi$ the boundary-curve could be expanded in
powers of a deformation-parameter $\epsilon$
\be
r(\psi) = \bar r (1+2\epsilon
\cos(p\psi)+ 2\sum_{n=2}^{\infty} c_n(\epsilon) \cos(np\psi)). \label{exp}
\ee
The coefficients $c_n=O(\epsilon^n)$ were determined up to order $\epsilon^7$.
Although one has solutions for $p-2$ different densities $\rho$, it turned out
that $r(\psi)$ was the same for all these densities.

Here a non-perturbative solution is given. It is shown that the
boundary-curve obeys the differential equation
\be
\frac 1{\sqrt{r^2+r^{\prime 2}}} = ar^2+b+cr^{-2}, \label{diffc}
\ee
with $r'=\de r/\de\psi$. This equation can be integrated
\bea
r^{\prime 2} &=& \frac {r^4}{(ar^4+br^2+c)^2}-r^2, \\
\psi &=& \pm\int \frac{\de r(ar^4+br^2+c)}{r\sqrt{r^2-(ar^4+br^2+c)^2}} \nn
&=& \pm\int^{r^2} \frac{\de q(aq^2+bq+c)}{2q\sqrt{q-(aq^2+bq+c)^2}} \label{psi}
\eea
and is thus given by an elliptic integral. With increasing $\psi$ the radius
$r$ oscillates periodically between the largest and the smallest radii $r_>$
and $r_<$, resp. The three constants $a$, $b$, and $c$ are determined by
these extreme radii, and by the periodicity of the boundary. Since $r'$
vanishes for the extrema of $r$, one has
\be
r_{>,<} = ar_{>,<}^4+br_{>,<}^2+c.
\ee
The periodicity is given by
\be
\int_{r_<}^{r_>} \frac{\de r(ar^4+br^2+c)}{r\sqrt{r^2-(ar^4+br^2+c)^2}}
= \frac{\pi}p.
\ee
The solution of the differential equation (\ref{diffc}) expanded in powers of
$\epsilon$ agrees completely with the expansion obtained in \cite{Wegner}.

I explain now, how I arrived at this differential equation.
Denote the two intersections of the water-line with the cross-section by $L$
and $R$. From the general arguments given
in \cite{Wegner} one knows, that the length $2l$ of the water-line has to be
independent of the orientation. (Here actually we will fix the orientation of
the cross-section and rotate the water-line including the direction of the
gravitational force.)  The midpoint $M$ of the line moves in the
direction of the chord $LR$ as the water-line is rotated. Suppose the
coordinates of the intersections are given by $L=(x_M-l,y_M)$ and
$R=(x_M+l,y_M)$. If the line moves by an infinitesimal amount $L'=(x_M-l+\de
x_L,y_M+\de y_L)$, $R'=(x_M+l+\de x_R,y_M+\de y_R)$, then the condition, that
the distance $L'R'=2l$ is fixed yields $\de x_L=\de x_R$ and the condition,
that $M$ moves in the direction of $LR$ yields $\de y_L+\de y_R=0$. From both
conditions we find, that under this change the same length of the perimeter
$\de u=\sqrt{\de x_L^2+\de y_L^2}=\sqrt{\de x_R^2+\de y_R^2}$
disappears on one side below the water-line and appears on the other side
above the water-line. Then $\de x_L=\de x_R=\cos\beta\de u$,
$\de y_L=-\de y_R=\sin\beta\de u$. Thus the angles between the tangents on the
boundary and the water-line are the same on both ends.

\begin{figure}[h]
\center{\epsfig{file=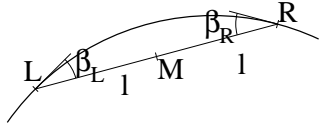,scale=0.5}}
\caption{Boundary curve, water-line $LR$, tangents, and the angles $\beta$.}
\end{figure}

In the following we will use these properties in two forms: (i) as
line-condition, requiring, that the envelope of the water-lines is the loci of
the midpoints $M$, so that we obtain $L$ and $R$ by drawing tangents on the
envelope and going a constant distance $l$ along the tangent in both directions
to obtain $L$ and $R$. (ii) as angle-condition, requiring, that moving $L$ and
$R$ the same piece $u$ along the perimeter to $L'$ and $R'$, the angles
$\beta_L$ and $\beta_R$ between the tangents and the chord $L'R'$
always obey $\beta_L=\beta_R$. Both conditions are equivalent. {From} (i)
follows (ii) and from (ii) follows (i).

In a next step consider the limit of
large $p$. Then $r$ will oscillate on a distance of order $\bar r/p$, which is
small in comparison to $\bar r$ itself. I assume, that then one can
neglect the overall average curvature $1/\bar r$ and one may find a periodic
function $f$ of $\xi=\bar r\psi$, which obeys the line-condition. Indeed
assuming, that this holds for $l\gg \bar r/p$, that is on a distance of many
periods, one can derive in this linear case a differential equation, which has
to be obeyed by the function $f$. This will be done in
subsection \ref{ssl1}. Expanding the solution in powers of $\epsilon$ is in
agreement with the terms leading in $1/p$ which were found in eq. 86 of
\cite{Wegner}.

Unfortunately this procedure is not applicable for finite $p$. Here I work
with a new conjecture: Although for physical reasons the periodicity $p$ has
to be an integer, one may try it with non-integer $p$. If $p$ differs from an
integer infinitesimally, then going around by the angle $2\pi$ one arrives at
the same function shifted however, in $\psi$-direction by an infinitesimal
angle $2\alpha$. I conjecture that there is also a line between $L$
and $R$ belonging to angles $\psi_{L,R}$ differing by $\delta\psi\approx 2\pi$
of constant small length $2l$, which obeys the line-condition. This condition
yields a differential equation. For the linear case this is done in
subsection \ref{ssl2}. It is in agreement with that derived for large
distance $l$. For the circular case one obtains the differential equation in
subsection \ref{ssc1}, which is identical to eq. (\ref{diffc}).

Finally it is shown in subsections \ref{ssl3} and \ref{ssc2}, that the
angle-condition holds for the curves described by the so obtained differential
equations which are shifted by arbitrary distances in the linear case and by
arbitrary angles in the circular case, resp. This last property can now also
be used for two points $L$ and $R$ on the same curve, and thus shows that eq.
(\ref{diffc}) yields the boundaries for cross-sections with $\rho\not=1/2$. 

The derivation for the linear approximation is given, since it
is somewhat simpler, but one may skip section \ref{secl} and
go directly to the derivation in section \ref{secc}.

\section{Linear Approximation\label{secl}}

Consider now the above mentioned linear case. Let us start from the loci of
the middle points $M$ of the chord $LR$, described by Cartesian coordinates
$(g,\xi)$. Then drawing
the tangent on this curve and moving along the tangent by the distance $l$ in
both directions one obtains the points $R$ and $L$, which should lie on the
same curve described by $f(\xi)$. Thus the functions $g$ and $f$ should obey
\be
g(\xi)\pm \frac{lg'(\xi)}{\sqrt{1+g^{\prime 2}(\xi)}}
= f(\xi \pm \frac l{\sqrt{1+g^{\prime 2}(\xi)}}).
\ee

\subsection{Large Distance \label{ssl1}}

\newcommand{\pil}{\lambda}
Assume now, that the functions are periodic with periodicity $2\pil$
and $f(\xi+\pil)=-f(\xi)=f(-\xi)$, similarly for $g$. Then $f(n\pil)=0$ for
integer $n$ and $f$ has extrema at $\xi=o\pil/2$ with odd $o$. If we seek
solutions for large $l$, then we put $l=l_o+\delta$ with
$l_o=o\pil/2$ and small $\delta$. We write
\be
\frac l{\sqrt{1+g^{\prime 2}(\xi)}} = l_o +\delta_1,
\quad \delta_1 = \frac{l_o+\delta}{\sqrt{1+g^{\prime 2}(\xi)}}-l_o
=\delta-l_o g^{\prime 2}/2 + ...
\ee
and obtain by expansion in $\delta_1$ and addition and subtraction of the
equations
\bea
g(\xi) &=& (-)^{(o-1)/2}(\delta_1 f'(\xi+\pil/2) + \frac{\delta_1^3}6
f^{\prime\prime\prime}(\xi+\pil/2)+...), \\
(l_o+\delta) \frac{g'(\xi)}{\sqrt{1+g^{\prime 2}(\xi)}}
&=& (-)^{(o-1)/2} (f(\xi+\pil/2) + \frac{\delta_1^2}2
f^{\prime\prime}(\xi+\pil/2)+...).
\eea
Note that $f(\xi+l_o+\delta_1)=(-)^{(o-1)/2}f(\xi+\frac{\pil}2+\delta_1)$ and
$f(\xi-l_o-\delta_1)=(-)^{(o+1)/2}f(\xi+\frac{\pil}2-\delta_1)$.
If $f$ is of order $l_o^0$, then we realize from the second equation, that
$g'$ and thus $g$ is of order $1/l_o$. The first equation yields, that
$\delta_1$ is of order $1/l_o$. We put $\delta=c/l_o$. Furthermore we write
$(-)^{(o-1)/2}l_og=\hat g$ and collect the leading terms of these two equations
\be
\hat g = (c-\hat g^{\prime 2}/2) f', \quad \hat g'=f,
\ee
where the argument of $g$ is $\xi$ and of $f$ is $\xi+\pil/2$. From these two
equations we obtain differential equations for $f$ and $\hat g$. Taking the
derivative of the second equation and substituting into the first one yields
\be
\hat g = (c-\hat g^{\prime 2}/2) \hat g^{\prime\prime}.
\ee
If we multiply this equation by $\hat g'$, we can integrate it and obtain
\be
\hat g^2 = (c-\hat g^{\prime 2}/4) \hat g^{\prime 2} +c_2.
\ee
In order to obtain a differential equation for $f$ we solve this equation for
$\hat g$ and differentiate it. This yields
\be
\hat g' = \frac{\hat g^{\prime\prime}(c\hat g'-\hat g^{\prime 3}/2)}
{\sqrt{c_2+c\hat g^{\prime 2}-\hat g^{\prime 4}/4}}
\ee
and substituting $\hat g'=f$ the equation
\be
\sqrt{c_2+cf^2-f^4/4}=(c-f^2/2)f'
\ee
or
\be
(c-f^2/2)^2(1+f^{\prime 2}) = c_2+c^2, \label{eq1}
\ee
which we will use in the form
\be
\frac 1{\sqrt{1+f^{\prime 2}}} = af^2+b
\label{diffl}
\ee
with new integration constants $a$ and $b$.

\subsection{Infinitesimal distance\label{ssl2}}

We perform now a similar calculation, but assume, that we can represent
$f(\xi+\alpha)$ and $f(\xi-\alpha)$ similarly by $g$, where now $\alpha$ and
$l$ are infinitesimally small. Thus we assume
\be
g(\xi)\pm \frac{lg'(\xi)}{\sqrt{1+g^{\prime 2}(\xi)}}
= f(\xi \mp \alpha \pm \frac l{\sqrt{1+g^{\prime 2}(\xi)}}).
\ee
It will turn out, that we need the expansion in $\alpha$ to first order and in
$l$ in third order. We obtain
\bea
g&=&f+\frac{l^2}{2(1+g^{\prime 2})}f^{\prime\prime}, \\
g'&=&f'-\frac{\alpha}l f'\sqrt{1+g^{\prime 2}}
+\frac{l^2}{6(1+g^{\prime 2})}f^{\prime\prime\prime}.
\eea
Differentiating the first equation yields
\be
g'=f'+\frac{l^2f^{\prime\prime\prime}}{2(1+g^{\prime 2})}
-\frac{l^2g'g^{\prime\prime}f^{\prime\prime}}{(1+g^{\prime 2})^2}.
\ee
Equating both expressions and choosing $\alpha=cl^3$ yields
\be
cf'\sqrt{1+g^{\prime 2}} 
+ \frac{f^{\prime\prime\prime}}{3(1+g^{\prime ^2})}
-\frac{g'g^{\prime\prime}f^{\prime\prime}}{(1+g^{\prime 2})^2} = 0.
\ee
In the limit $\alpha\rightarrow 0$ one has $g=f$. With this substitution and
multiplication by $1/\sqrt{1+f^{\prime 2}}$ we may integrate this equation
\be
cf+\frac{f^{\prime\prime}}{3(1+f^{\prime 2})^{3/2}}+c_1=0.
\ee
Multiplication by $f'$ allows a second integration which yields
\be
\frac 1{\sqrt{1+f^{\prime 2}}}=\frac{3c}2f^2+3c_1f+3c_2. \label{diffs}
\ee
Comparison with the equation obtained for large distances (\ref{diffl}) shows
agreement, if we change the notation for the constants. There is no linear
term in eq.(\ref{diffl}) like the term $3c_1f$ in eq.(\ref{diffs}). The reason
is, that in eq. (\ref{diffl}) we required $f$ to vanish on the average. But of
course a constant can be added to $f$ and $g$, yielding such a linear term.
Thus both procedures yield the same differential equation.

\subsection{Arbitrary Distance\label{ssl3}}

We show now the following {\bf theorem}. Consider two curves $C_L$ and $C_R$
which are governed by eq. (\ref{diffl}) with the same constants $a$ and $b$,
and point $L\in C_L$ with Cartesian coordinates $\xi_L,f_L$ and
$R\in C_R$ with coordinates $\xi_R,f_R$. If now the angle
$\beta_L$ between the tangent in $L$ and $\overline{LR}$ equals
the angle $\beta_R$ between the tangent in $R$ and $\overline{RL}$, and
$f_L\not=f_R$, then this property will hold between any two points $L'\in
C_L$ and $R'\in C_R$, which are obtained by moving from $L$ and $R$ by
the same arc $u$ in the same direction along the curves. Simultaneously the
distance $\overline{L'R'}=\overline{LR}$ is constant.

In order to show this we first express the angles $\beta_L$ and $\beta_R$. Let
us denote the angles of the curves against the $\xi$-axis by $\phi_{L,R}$,
\bea
\tan\phi_{L,R} &=& f'_{L,R}, \quad 
\cos\phi_{L,R}=\frac 1{\sqrt{1+f_{L,R}^{\prime 2}}}=af_{L,R}^2+b, \\
\sin\phi_{L,R}&=&\frac{f'_{L,R}}{\sqrt{1+f_{L,R}^{\prime 2}}}
=\sqrt{1-(af_{L,R}^2+b)^2}=:W_{L,R}
\eea
and the angle of $LR$ against the $\xi$-axis by $\gamma$,
\be
\tan\gamma=\frac{f_R-f_L}{\xi_R-\xi_L}, \quad
\sin\gamma=\frac{f_R-f_L}{2l}, \quad
\cos\gamma=\sqrt{1-\frac{(f_R-f_L)^2}{(2l)^2}}=:W_{\gamma},
\ee
where $2l$ is the distance $LR$.

\begin{figure}[h]
\center{\epsfig{file=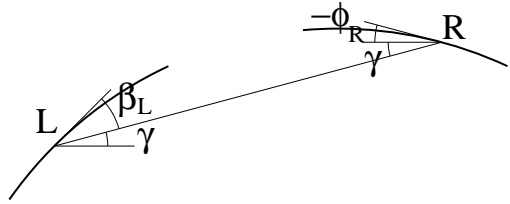,scale=0.8}}
\caption{In this figure $\phi_L$ is given by $\beta_L+\gamma$. The
angle $\phi_R$ is negative in this figure, and $\beta_R=\gamma-\phi_R$.}
\end{figure}

Then we have $\beta_L=\phi_L-\gamma$, $\beta_R=\gamma-\phi_R$. The condition
$\beta_L=\beta_R$ can now be expressed in various ways. From
$\cos(\phi_L-\gamma)=\cos(\gamma-\phi_R)$ one obtains
\be
(af_L^2+b)W_{\gamma}+\frac{f_R-f_L}{2l}W_L
=(af_R^2+b)W_{\gamma}+\frac{f_R-f_L}{2l}W_R,
\ee
which for $f_R\not= f_L$ reduces to
\be
a(f_R+f_L)W_{\gamma} = \frac{W_L-W_R}{2l}. \label{cl1}
\ee
We show now, that
\be
n:=\beta_L-\beta_R = \phi_L+\phi_R-2\gamma
\ee
is independent of $u$, if initially it vanishes, $n=0$, and
$f_R\not=f_L$ holds. For this purpose we calculate the derivative
$\frac{\de n}{\de u}$ and show, that it vanishes, and thus, that $n$ remains 0.
For this purpose we calculate the derivatives needed and indicate the
derivative with respect to u by a dot,
\bea
\dot f_{L,R} &=& \sin(\phi_{L,R})=W_{L,R}, \\
\dot W_{L,R} &=& \frac{\de W_{L,R}}{\de f_{L,R}}\dot f_{L,R}
=-2af_{L,R}(af_{L,R}^2+b), \\
\dot W_{\gamma} &=& -\frac{(f_L-f_R)(W_L-W_R)}{4l^2W_{\gamma}}.
\eea
We use generally for the derivatives of the angles
\be
\dot\phi = \cos\phi(\sin\phi\dot) -\sin\phi(\cos\phi\dot)
\label{diffang}
\ee
and obtain
\be
\dot\phi_{L,R} = -2af_{L,R}
\ee
and
\be
\dot\gamma = W_{\gamma}\frac{W_R-W_L}{2l}
-\frac{(f_R-f_L)^2(W_L-W_R)}{8l^3W_{\gamma}}.
\ee
We make now use of (\ref{cl1}) and obtain
\be
\dot\gamma=-a(f_L+f_R),
\ee
from which we conclude $\dot n=0$, provided $f_L\not=f_R$ \rule{2mm}{2mm}

For $f_L=f_R$ one has $W_L=\pm W_R$. If the minus-sign applies, then $n$
vanishes for all $f_L$ and $l$. In this case, however, $W_{\gamma}=1$, 
$\dot W_{\gamma}=0$ and we obtain $\dot n=-4af+2W_L/l$. One can easily check,
that for $f_R=f_L+\epsilon$ one obtains
$n=\epsilon(2af/W_L-1/l)+O(\epsilon^2)$, so that $\dot n=0$ holds also
as we approach $f_L=f_R$. Thus the theorem is proven.

Obviously instead of considering two curves, we may consider two points $L$
and $R$ on the same curve, which implies, that the midpoints of these chords
are the envelope of these lines as we increase $u$.

\section{Circular Case\label{secc}}

The point $R$ with polar coordinates $(r,\psi+\delta\psi)$ lie on the tangent
of the envelope through $M$ with polar coordinates $(\mu,\psi)$ a distance $l$
apart. Then we have
\be
r^2 = \mu^2+\frac{2\mu\mu'l}{\sqrt{\mu^2+\mu^{\prime
2}}}+l^2
\ee
Denote the angle $OMR$ by $\gamma$. Then $\gamma$ is related to the
slope of the tangent by $-\cos\gamma=\mu'/\sqrt{\mu^2+\mu^{\prime 2}}$ and
$\sin\gamma=\mu/\sqrt{\mu^2+\mu^{\prime 2}}$. For $\delta\psi$ we obtain \\
$\tan\delta\psi=l\sin\gamma/(\mu-l\cos\gamma)=l\mu/(\mu\sqrt{\mu^2+\mu^{\prime
2}}+l\mu')$.
Thus we obtain
\be
\sqrt{\mu^2\pm\frac{2\mu\mu'l}{\sqrt{\mu^2+\mu^{\prime2}}}+l^2}
=r(\psi\mp\alpha\pm\arctan\frac{l\mu}{\mu\sqrt{\mu^2+\mu^{\prime 2}}\pm
l\mu'}), \label{mtor}
\ee
where the upper sign yields $r_R$ on the curve rotated by $\alpha$ and the
lower sign $r_L$ on the curve rotated by $-\alpha$.

\begin{figure}[ht]
\center{\epsfig{file=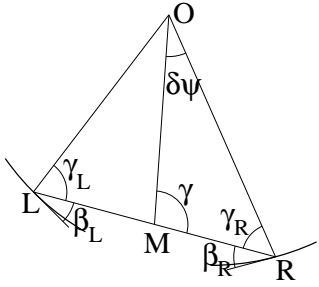,scale=0.5}}
\caption{The angles in the circular case. The angles $\phi$ determined by
$\tan\phi=r'/r$ are related to $\beta$ and $\gamma$ by
$\phi_L=\gamma_L+\beta_L-\pi/2$, $-\phi_R=\gamma_R+\beta_R-\pi/2$.}
\end{figure}

\subsection{Infinitesimal angle\label{ssc1}}

We expand both sides of eq. (\ref{mtor}) up to order $l^3$ and $\alpha$ and add
and subtract the equations for both signs. Then we obtain
\bea
\mu+\frac{l^2\mu}{2(\mu^2+\mu^{\prime 2})}
&=& r + l^2 (-\frac{r'\mu'}{\mu(\mu^2+\mu^{\prime 2})} 
-\frac{r^{\prime\prime}}{2(\mu^2+\mu^{\prime 2})}), \label{mtor2}\\
\mu'-\frac{l^2\mu'}{2(\mu^2+\mu^{\prime 2})}
&=& r' -\frac{\alpha}l r'\sqrt{\mu^2+\mu^{\prime 2}}
+l^2(\frac{r'(-\mu^2+3\mu^{\prime 2})}{3\mu^2(\mu^2+\mu^{\prime 2})} \nn
&&-\frac{r^{\prime\prime}\mu'}{\mu(\mu^2+\mu^{\prime 2})}
+\frac{r^{\prime\prime\prime}}{6(\mu^2+\mu^{\prime 2})}). \label{mtor3}
\eea

Now we differentiate eq. (\ref{mtor2})
\bea
&&\mu'+ l^2(\frac{\mu'(-\mu^2+\mu^{\prime 2)}}{2(\mu^2+\mu^{\prime 2})^2} 
- \frac{\mu\mu'\mu^{\prime\prime}}{(\mu^2+\mu^{\prime 2})^2}) \nn
&=& r' 
+l^2(\frac{r'\mu^{\prime 2}(3\mu^2+\mu^{\prime 2})}{\mu^2(\mu^2+\mu^{\prime
2})^2} 
-\frac{r^{\prime\prime}\mu'(2\mu^2+\mu^{\prime 2})}{\mu(\mu^2+\mu^{\prime
2})^2} \nn
&&+\frac{\mu^{\prime\prime}r'(-\mu^2+\mu^{\prime 2})}{\mu(\mu^2+\mu^{\prime
2})^2} 
- \frac{r^{\prime\prime}\mu^{\prime\prime}\mu'}{(\mu^2+\mu^{\prime^2})^2}
+ \frac{r^{\prime\prime\prime}}{2(\mu^2+\mu^{\prime 2})})
\eea
and subtract eq. (\ref{mtor3}) from this derivative. The resulting equation
contains terms proportional to $\alpha/l$ and to $l^2$. Thus we set $\alpha$
proportional $l^3$. In the limit $l\rightarrow 0$ the function $\mu$ approaches
$r$ and similarly for the derivatives. Thus we have
\bea
f:&=&-\frac{\alpha}{l^3} r'\sqrt{r^2+r^{\prime 2}} \nn
&-&\frac{r'(r^2+4r^{\prime 2})}{3(r^2+r^{\prime 2})^2}
+\frac{r'r^{\prime\prime}(r^2-r^{\prime 2})}{r(r^2+r^{\prime 2})^2}
+\frac{r'r^{\prime\prime 2}}{(r^2+r^{\prime 2})^2}
-\frac{r^{\prime\prime\prime}}{3(r^2+r^{\prime 2})} = 0.
\eea

This differential equation can be integrated. One finds, that
\be
f= \frac{\sqrt{r^2+r^{\prime 2}}}r \frac{\de f_1}{\de \psi}
= \frac{\sqrt{r^2+r^{\prime 2}}}{r^3} \frac{\de f_3}{\de \psi}
\ee
with
\bea
f_1 &=& \frac{-r^{\prime\prime}r+r^2+2r^{\prime 2}}{3(r^2+r^{\prime 2})^{3/2}}
- \frac{\alpha r^2}{2l^3}, \\
f_3 &=& -\frac{r^3(r+r^{\prime\prime})}{3(r^2+r^{\prime 2})^{3/2}}
-\frac{\alpha r^4}{4l^3},
\eea

Thus $f_1$ and $f_3$ are constants. Elimination of $r^{\prime\prime}$ yields
the differential equation of the bounding curve quoted in the introduction,
eq. (\ref{diffc}).
\be
\frac 1{\sqrt{r^2+r^{\prime 2}}} = ar^2 + b + cr^{-2}
\ee
with constants $a=\lim 3\alpha/l^3$, $b=3f_1/2$, and $c=-3f_2/2$.

\subsection{Arbitrary angle\label{ssc2}}

Now we consider the two curves rotated against each other by an arbitrary
angle $2\alpha$ and perform similar calculations to those for the linear case.

We first express the angles
\be
\beta_L=\phi_L-\gamma_L+\frac{\pi}2, \quad
\beta_R=\frac{\pi}2-\phi_R-\gamma_R.
\ee
and show the following {\bf theorem}: If the condition $\beta_L=\beta_R$, that
is \be
n:=\phi_L-\gamma_L+\phi_R+\gamma_R=0
\ee
is fulfilled and $r_L\not=r_R$ and the points $L$ and $R$ move along their
curves by the same arc $u$ to points $L'$ and $R'$, then the condition
$n=0$ remains fulfilled, that is for all these chords $\beta_L=\beta_R$ holds.
Simultaneously the distance $\overline{LR}=\overline{L'R'}$ remains constant.

In order to show this we calculate $\de n/\de u$. For this purpose we first
list the sines and cosines of the angles
\bea
\cos\phi_{L,R} &=& \frac{r_{L,R}}{\sqrt{r_{L,R}^2+r_{L,R}^{\prime 2}}} 
= r_{L,R}(ar_{L,R}^2+b+\frac c{r_{L,R}^2}) \\
\sin\phi_{L,R} &=& \frac{r'_{L,R}}{\sqrt{r_{L,R}^2+r_{L,R}^{\prime 2}}}
=:W_{L,R} \\
\cos\gamma_{L,R} &=& \frac{4l^2+r_{L,R}^2-r_{R,L}^2}{4lr_{L,R}} \\
\sin\gamma_{L,R} &=& \frac{W_{\gamma}}{4lr_{L,R}} \\
W_{\gamma} &=& \sqrt{-16l^4+8l^2(r_L^2+r_R^2)-(r_L^2-r_R^2)^2}
\eea
Their derivatives with respect to $u$ indicated by a dot are
\bea
\dot r_{L,R} &=& \sin\phi_{L,R} = W_{L,R} \label{dotr} \\
\dot W_{L,R} &=& -\cos\phi_{L,R} \frac{\de\cos\phi_{L,R}}{\de r_{L,R}}
\eea
Performing the derivatives for $r_L$ and $r_R$ and using (\ref{dotr}) one
obtains
\be
\dot W_{\gamma} = \frac{8r_Lr_Rl}{W_\gamma}(W_R\cos\gamma_L+W_L\cos\gamma_R)
\label{dotg}
\ee
Thus we obtain using again eq. (\ref{diffang})
\bea
\dot\phi_{L,R} &=& -\cos^2\phi_{L,R} \frac{\de \cos\phi_{L,R}}{\de r_{L,R}}
-W_{L,R}^2 \frac{\de \cos\phi_{L,R}}{\de r_{L,R}} \nn 
&=& -\frac{\de \cos\phi_{L,R}}{\de r_{L,R}} 
= -3ar_{L,R}^2 -b + \frac c{r_{L,R}^2}, \\
\dot\gamma_{L,R} &=& \frac{\cos\gamma_{L,R}\dot W_{\gamma}}{4lr_{L,R}}
+\frac{W_{\gamma}(r_{R,L}W_{R,L} -r_{L,R}W_{L,R})}{8l^2r_{L,R}^2}.
\eea
Now we use $n=0$ in order to simplify the derivatives $\dot\gamma$.
First we evaluate $\sin(\phi_L+\gamma_R)=\sin(\gamma_L-\phi_R)$ which yields
\be
W_R\cos\gamma_L+W_L\cos\gamma_R 
= \frac{W_{\gamma}}{4l}(\frac{\cos\phi_R}{r_L}-\frac{\cos\phi_L}{r_R})
\label{wgamma}
\ee
Inserting this expression into eq.(\ref{dotg}) yields
\be
\dot W_{\gamma} =2(r_R\cos\phi_R-r_L\cos\phi_L)
\ee
Further we determine the products $W_{\gamma}W_{L,R}$ from
$\cos(\phi_L-\gamma_L)=\cos(\phi_R+\gamma_R)$ and 
$\cos(\phi_L+\gamma_R)=\cos(\phi_R-\gamma_L)$, which yield
\bea
\cos\phi_L\cos\gamma_L + \frac{W_{\gamma}W_L}{4lr_L}
&=& \cos\phi_R\cos\gamma_R - \frac{W_{\gamma}W_R}{4lr_R}, \\
\cos\phi_L\cos\gamma_R - \frac{W_{\gamma}W_L}{4lr_R}
&=& \cos\phi_R\cos\gamma_L + \frac{W_{\gamma}W_R}{4lr_L}.
\eea
For $r_L\not= r_R$ one obtains
\bea
W_{\gamma}W_L &=& 8al^2r_Lr_R^2
+ (ar_L^3+br_L)(4l^2+r_R^2-r_L^2)
-\frac {c(4l^2+r_L^2-r_R^2)}{r_L}, \\
W_{\gamma}W_R &=& -8al^2r_Rr_L^2
- (ar_R^3+br_R)(4l^2+r_L^2-r_R^2)
+\frac {c(4l^2+r_R^2-r_L^2)}{r_R}
\eea
and thus
\bea
\dot\gamma_L &=& -a(r_L^2+2r_R^2) -b +\frac c{r_L^2}, \\
\dot\gamma_R &=& a(2r_L^2+r_R^2) +b - \frac c{r_R^2},
\eea
from which one concludes
$\dot n = \dot\phi_L - \dot\gamma_L +\dot\phi_R +\dot\gamma_R = 0$ 
provided $r_L\not=r_R$ \rule{2mm}{2mm}

\newcommand{\nicht}[1]{}
\nicht{
Using these derivatives and the expression (\ref{diffang}) for the derivative
of the angles one obtains a quadratic polynomial in the $W$s
with coefficients rational in $r_R$ and $r_L$. Next
$W_{\gamma}$ is expressed by $W_L$ and $W_R$ by means of eq.(\ref{wgamma}).
Then we replace $W_LW_R$ by a rational function in $r_L$ and $r_R$ by using,
that $\cos(\phi_L+\phi_R)=\cos(\gamma_L-\gamma_R)$ yields
\be
W_LW_R = \cos\phi_L\cos\phi_R-\cos\gamma_L\cos\gamma_R
-\frac{W_{\gamma}^2}{16l^2r_Rr_L} \label{wlr}
\ee
and express the squares of the $W$s by there rational functions. Then we obtain
\be
\dot n=\frac{(r_L^2+r_R^2) D}{32l^4r_L^2r_R^2(ar_L^2+ar_R^2+b)}
\ee
with
\bea
D &=& -r_L^4 -r_R^4 -16l^4
+2r_L^2r_R^2 +8l^2r_L^2+8l^2r_R^2 \nn
&+&16a^2l^2r_L^2r_R^2(-r_L^2r_R^2-4r_L^2l^2-4r_R^2l^2) 
+16abr_L^2r_R^2l^2(-r_L^2-r_R^2-4l^2) \nn
&+&16acl^2(-r_L^4-r_R^4+4r_L^2l^2+4r_R^2l^2) -16b^2r_L^2r_R^2l^2 \nn
&+&16bcl^2(-r_L^2-r_R^2+4l^2)
-16c^2l^2.
\eea
On the other hand one obtains from eq.(\ref{wlr})
\be
(W_LW_R)^2 - \left(\cos\phi_L\cos\phi_R-\cos\gamma_L\cos\gamma_R
-\frac{W_{\gamma}^2}{16l^2r_Rr_L}\right)^2
=\frac{(r_L^2-r_R^2)^2 D}{64l^4r_L^2r_R^2} = 0
\ee
Thus $D$ vanishes for $r_L\not=r_R$, and $\dot n=0$.
}

For $r_L=r_R$ one has $\gamma_L=\gamma_R$ and $\phi_L=\pm\phi_R$. If the minus
sign applies, then $\beta_L=\beta_R$ and $n=0$. In this case we are not
allowed to divide by $r_L-r_R$, which we did in our derivation. It turns out,
however, that in this limit $\dot W_{\gamma}=0$ and one obtains
\bea
\dot n &=& \frac{W_{\gamma} W_L}{2r_Ll^2} +c_0(r_L), \label{rl=rr} \\
c_0(r_L) &=& -6ar_L^2 -2b + \frac {2c}{r_L^2} \nn
&=& \frac{2a}{r_L^2}((r_>^2+r_L^2)(r_<^2+r_L^2)-4r_L^4)
+\frac{2(r_>r_<-r_L^2)}{r_L^2(r_>+r_<)}. 
\eea
\nicht{This expression multiplied by $W_{\gamma} W_L/(2r_Ll^2)-c_0$ equals
$D(r_R=r_L)/(4r_L^2l^4)$. Thus if $\dot n=0$ in this case, also $D=0$ holds.}

\subsection{Application to the boundary}

We may now apply the theorem to our original problem. In this problem the two
curves are identical and constitute the boundary. Suppose we have now a curve
obeying the differential equation (\ref{diffc}) with integer $p$. Then we have
to look for points $L,R$ on this curve which have either the properties $n=0$
and $r_L\not=r_R$ or otherwise the properties $r_L=r_R$ and $\dot n=0$, eq.
(\ref{rl=rr}). We will now show, that there are at least $p-2$ solutions,
which obey the second condition. For this purpose we divide the curve into
pieces $A_1$, ... $A_p$ in which $r$ decreases from $r_>$ to $r_<$ as $\psi$
increases and into pieces $B_1$, ... $B_p$, where $r$ increases from $r_<$ to
$r_>$. We now look for solutions $r_L=r_R$ and $\dot n=0$, where $r_R$ is in
$A_1$ and $r_L$ in $B_k$. Thus we have to find a solution of (\ref{rl=rr}). We
realize, that with the exception of $l=0$ the first term in (\ref{rl=rr})
vanishes for $r=r_>$ and $r=r_<$, since $W_L=0$, because $r'=0$ at these
extrema.

\begin{figure}[ht]
\center{\epsfig{file=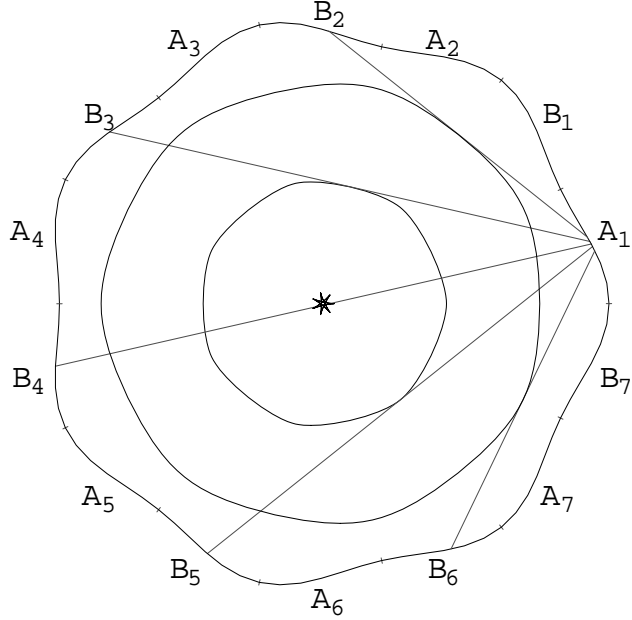,angle=270,scale=0.5}}
\caption{Boundary for $p=7$ and $\epsilon=1/49$ (eq. \ref{exp}) with the pieces
$A_n$ and $B_n$. The outermost curve is the boundary, the inner curves are the
envelopes of the water-lines. The innermost one for density $\rho=1/2$, the
two others for densities 0.22748 and 0.77252, and for 0.05269 and 0.94731.}
\end{figure}

On the other hand we can evaluate $c_0$ for these extrema
\bea
c_0(r_>) &=& -\frac{2(r_>-r_<)(1+2ar_>(r_>+r_<)^2)}{r_>(r_>+r_<)} \\
c_0(r_<) &=& \frac{2(r_>-r_<)(1+2ar_<(r_>+r_<)^2)}{r_<(r_>+r_<)}
\eea
The denominator $\sqrt{r^2-(ar^4+br^2+c)^2}$ of the integral (\ref{psi})
vanishes only at the extreme points $r=r_>$ and $r=r_<$. If we express $b$ and
$c$ in terms of $a$, $r_>$, and $r_<$, then
\bea
r^2-(ar^4+br^2+c)^2 &=& (r_>-r)(r-r_<)
\left(\frac 1{r_>+r_<} + a(r+r_>)(r+r_<)\right) \nn
&\times& (r_>+r)(r+r_<)
\left(\frac 1{r_>+r_<} + a(r-r_>)(r-r_<)\right) \nn
\eea
Thus $1+a(r+r_>)(r+r_<)(r_>+r_<) > 0$, and therefor the factors
$(1+2ar_>(r_>+r_<)^2)$ and $(1+2ar_<(r_>+r_<)^2)$ in the expressions for $c_0$
are positive. Therefore one has $c_0(r_>)<0$ and $c_0(r_<)>0$. Therefore
$\dot n$ changes sign between $r=r_>$ and $r=r_<$. Since $\dot n$ is a
continuous function of $r_L$, there must be a zero in between. This argument
holds for all pieces $B_k$ with the exception of those adjacent to $A_1$. (For
those adjacent $l$ may become 0. In this limit one finds $\dot n=0$.) Therefore
we have at least $p-2$ solutions, which is in agreement with those found in
\cite{Wegner}. We leave aside the question, whether for sufficiently large
ratios $r_>/r_<$ more solutions appear and whether there are other solutions
for $\rho\not=1/2$. We finally mention that in the near-circular case 
$(r_>-r_<)/(r_>+r_<)\ll 1$ the periodicity $p$ is related to $a$ by
$p^2 = 1+8ar^3$, where $r$ is some intermediate radius.

\section{Conclusion}

We have given a closed representation of the boundaries of logs floating in
any direction with $\rho\not=1/2$ in terms of a differential equation which
can be solved by an elliptic integral. The differential equation could be
obtained, since points of two of these curves rotated against each other can
be connected by chords so that by progressing on the two curves by the same
length of the perimeter the length of the chord does not change. Using that
this holds also for two curves rotated against each other by an infinitesimal
angle connected by a chord of infinitesimal length the differential equation
could be derived. The differential equation of the linear analogue of this
curve has also been given.

\end{document}